\documentclass{aa}  
\usepackage{graphicx}
\usepackage{txfonts}
\usepackage{amssymb}
\usepackage{natbib}
\bibpunct{(}{)}{;}{a}{}{,}
\newcommand{\pas}{.\hskip-2pt$^{\prime\prime}$}
\def\kms{km~s$^{-1}\,$}

\titlerunning{H$_2$O and CH$_3$OH maser associations in AFGL~5142 and Sh~2-255 IR}
\begin{document}
   \title{Associations of H$_2$O and CH$_3$OH masers at milli-arcsec angular resolution  in two high-mass YSOs}

  \author{ C. Goddi \inst{1}
\and L. Moscadelli \inst{2}
\and A. Sanna  \inst{2}
\and R. Cesaroni \inst{1}
\and V. Minier \inst{3}}
   \offprints{C. Goddi,\\\email{cgoddi@arcetri.astro.it}}

   \institute{ INAF, Osservatorio Astrofisico di Arcetri, Largo E. Fermi 5, 50125 Firenze, Italy
    \and INAF, Osservatorio Astronomico di Cagliari, Loc. Poggio dei Pini,
 Str. 54, 09012 Capoterra (CA), Italy   \and Service d'Astrophysique, DAPNIA/DSM/CEA Saclay, 91191 Gif-sur-Yvette, France }

 
  \abstract{
Most previous high-angular ($<0.1$ arcsec) resolution studies of molecular masers in high-mass star forming regions (SFRs) have concentrated mainly on either water or methanol masers. While high-angular resolution observations have clarified that water masers originate from shocks associated with protostellar jets, different environments have been proposed in several sources to explain the origin of methanol masers.
Tha aim of the paper is to investigate the nature of the methanol maser birthplace in SFRs and the association between the water and methanol maser emission in the same young stellar object.
We have  conducted phase-reference Very Long Baseline Interferometry (VLBI) observations of water and methanol masers toward two high-mass SFRs,  Sh 2-255 IR and AFGL 5142.
In Sh 2-255 IR water masers are aligned along a direction close to the orientation of the molecular outflow observed on angular scales of 1-10 arcsec, tracing possibly the disk-wind emerging from the disk atmosphere. 
In AFGL 5142 water masers trace expansion at the base of a protostellar jet, whilst  methanol masers are more probably tracing infalling than outflowing gas.
The results for AFGL 5142 suggest that water and methanol masers trace different kinematic structures in the circumstellar gas.

}

   \maketitle
%

\section{Introduction}

Surveys of  6.7~GHz CH$_3$OH and 22.2~GHz H$_2$O masers in high-mass star forming regions (SFRs) have shown that both maser species trace the earliest evolutionary stages of the high-mass star forming process \citep{Cod00,Beu02a}. 
Both maser types are often spatially coincident with massive molecular outflows and hot molecular cores, and occasionally with young HII regions \citep{Cod97,Wal98}. \citet{Beu02a} found that every source, where methanol maser emission  is associated with a radio continuum source, shows also water maser emission and suggest that methanol masers may disappear at an earlier evolutionary stage than H$_2$O masers.  In fact, it has been found that the 6.7~GHz  methanol masers are most likely present before an observable UC HII region is formed around a massive star and are quickly destroyed as the UC HII region evolves \citep{Wal98,Cod00}. 
In addition, single-dish  data revealed that the velocity spread of maser emission is usually larger for H$_2$O than for CH$_3$OH masers \citep{Sri02}, suggesting that the two maser species might form in different components of the circumstellar gas.

 Very Long Baseline Interferometry (VLBI) multi-epoch observations have clarified that water masers originate from shocks  associated with winds and/or jets
 ejected from the massive YSOs (e.g. \citealt{God05,Mos05}). 
High-angular resolution studies indicate a clear relationship between methanol masers and 
hot cores (e.g., G29.95-0.02), binary high-mass protostellar cores seen in IR (e.g., Sh~2-255 IR), 
and HII regions (e.g., IRAS20126+4104) \citep{Min01}.
In some cases, high-resolution imaging has shown that  CH$_3$OH sources (e.g., G309.92+0.48 and NGC7538) have maser spots 
in linear or arc-like structures 
with velocity gradients compatible with   Keplerian rotation, suggesting that these masers may originate 
in the innermost portions of a  disk rotating about the protostar   \citep{Nor98,Min00,Pes04,Edr05}.
 In other cases (e.g., G313.77-0.86), the 6.7~GHz spots are distributed parallel to the axis of the jet/outflow as traced by the shocked H$_{2}$ emission at 2.12$\,\mu$m   and it has been proposed that the methanol maser emission may trace the shock at the interface between the jet and the circumstellar molecular gas \citep{Deb03}.

 Notwithstanding the plethora of H$_2$O and CH$_3$OH maser studies, so far only few of these (e.g., \citealt{Edr05}) have been performed in {\it both} maser types with sufficient angular resolution  ($\leq$0\pas1) to investigate a possible association between  H$_2$O and CH$_3$OH masers with the {\it same}   young stellar object (YSO).
To investigate the nature of the maser birthplaces and to test the importance of maser observations as indicators of the evolutionary stage of high-mass YSOs, it is essential to conduct VLBI studies on SFRs where {\it both} types of masers are {\it truly} associated with the same YSO. To achieve this goal, the observations have to be performed in phase-reference mode,  to establish the absolute position of maser emission with an accuracy of a few milliarcseconds.

In the present paper, we present VLBI data of  6.7~GHz  CH$_3$OH and 22.2~GHz H$_2$O masers  toward two high-mass YSOs: Sh~2-255 IR and AFGL~5142. At a distance of 2.5~kpc \citep{Eva77} and 1.8~kpc \citep{Sne88}, the bolometric luminosity of the two SFRs are estimated to be $ 9 \times 10^3$  L$_\odot$ \citep{Mez90} and $4 \times 10^3$ L$_\odot$ \citep{Hun99}, respectively. 
Both Sh~2-255 IR and AFGL~5142 are the reddest sources in a cluster of forming stars observed in the NIR continuum  emission, are associated with a young HII region, and present both methanol and water maser emission, indicating that very likely they share a common evolutionary (possibly zero age main sequence -- ZAMS) stage.
The 22.2~GHz H$_2$O masers in AFGL~5142 were previously  observed with the  Very Long Baseline Array (VLBA) by \citet{God06} \defcitealias{God06}{Paper~I} (hereafter, Paper I). Here, we present new (single-epoch) European VLBI Network (EVN) observations of the  6.7~GHz  CH$_3$OH masers.
 In Sh~2-255 IR, both CH$_3$OH  and  H$_2$O  masers were previously   observed with the European VLBI Network (EVN)   by \citet{Min00} and \citet{God05}, respectively. Owing to  the limited sensitivity  (average detection threshold of $\sim$0.3~Jy~beam$^{-1}$) and the too long  time separation  between consecutive epochs ($\geq$ 3~months) of the 22~GHz observations, only the  relative proper motions of  the strongest and more-longeval maser features were measured by \citet{God05}. 
Hence, we have performed follow-up observations of the 22~GHz H$_2$O masers in  Sh~2-255 IR  using the VLBA, taking advantage of both  high sensitivity and shorter  time separation  between  consecutive epochs. The VLBA observations were performed in phase-reference mode to derive the absolute positions and absolute proper motions of the water maser features.

Section~\ref{obs} describes our  multi-epoch VLBA and single-epoch EVN observations and gives technical details on the data reduction. 
Section~\ref{res} compares our VLBI maps of  CH$_3$OH and H$_2$O masers   with previous interferometrical observations of the SFRs Sh~2-255 IR and AFGL~5142.
Section~\ref{var} discusses the variability of both CH$_3$OH and H$_2$O maser emission based on single-dish monitoring.
In Sect.~\ref{afgl_dis} we analyze the relationship between   CH$_3$OH and H$_2$O masers in AFGL~5142, in particular investigating plausible birthplaces for 
CH$_3$OH masers. 
In Sect.~\ref{s255_dis} are discussed plausible kinematic scenarios for interpreting positions and proper motions of water maser features in Sh~2-255 IR. 
Conclusions are drawn in Sect.~7.


\section{Observations and data reduction}
\label{obs}

\subsection{VLBA observations}
Sh~2-255 IR was observed  in the \(6_{16}-5_{23}\) H$_2$O maser line (rest frequency 
22235.080~MHz) using the VLBA at four epochs (21 October and 08 December 2004, 21 January and 10 March 2005), each epoch lasting for 12 hours.  The nominal maser position used at the correlator was  \ $\alpha(J2000) = 06^h 12^m 54^s$.018,   $\delta(J2000) = 17^{\circ} 59' 23$\pas22.
The  observations were performed in phase-reference mode, alternating scans on   the maser source, Sh~2-255 IR, and the phase-reference source, J0603+1742, with a  switching  cycle of 70~s. J0603+1742 is separated 2.3$^{\circ}$  from Sh~2-255 IR, belongs to the VLBA calibrator catalog, and has very accurate coordinates (R.A. and Dec uncertainties $< 1 $~mas). Interposed every $\approx$80~min, 3--min scans on several continuum sources (J0449+1121, J0530+1331, J0555+3948, J0725+1425) were observed  for calibration purposes. 

Both circular polarizations were recorded using a 16~MHz bandwidth centered on the  line-of-sight (L.O.S.) velocity of 10~km~s$^{-1}$. The data were correlated with the VLBA FX correlator in Socorro (New Mexico) with an integration period of 2~s. The correlator used  1024 spectral channels corresponding to a channel separation of 0.2~km~s$^{-1}$.

\subsection{EVN observations}
AFGL~5142 was observed in the  5$_1\to6_0$ A$^+$ line of  methanol at 6.668~GHz using the EVN  on November 4 and 5, 2004, for a total of 12 hours. The antennae involved in the observations were Medicina, Cambridge, Onsala, Effelsberg,  Noto,  Darnhall, Westerbork, and Torun.
The  observations were performed in phase-reference mode, alternating scans on   the maser source, AFGL~5142, and the phase-reference source, J0518+3306, with a  switching cycle of 3~min.
 J05181+33062 is separated 2.6$^{\circ}$  from AFGL~5142, belongs to the VLBA calibrator catalog, and has very accurate coordinates (R.A. and Dec uncertainties $< 1 $~mas). For the purpose of bandpass and phase calibration, a scan of a few minutes on the strong, compact calibrator DA193 was observed every 1.5 h. The total on-source integration time for the maser target was about 5.5 hours.
Both circular polarizations were recorded with a 2~MHz bandwidth centered on the  L.O.S. velocity of 2~km~s$^{-1}$. 
The data were processed with the EVN MKIV correlator at the Joint Institute  for VLBI in Europe (JIVE -- Dwingeloo, The Netherlands),  with an integration period of 2~s, obtaining 1024 spectral channels with a separation of 0.088~km~s$^{-1}$.

\subsection{Data Reduction}
Data reduction was performed using the  NRAO's Astronomical Image Processing System (AIPS) package, following the standard procedure for VLBI line data.
Total power spectra of the continuum calibrators were used to derive the bandpass response of each antenna. The amplitude calibration was performed using the information on the system temperature and the gain curve of each antenna. For the Cambridge and Darnhall antennae  no information on the system temperatures were available, so for such antennae the  amplitude calibration was performed using the ``template spectrum'' method (which consists of comparing total power spectra of different scans).
For each source and observing epoch, a single scan of a strong calibrator was used to derive the instrumental (time-independent) single-band delay  and the phase offset between the two polarizations.
After removing the instrumental errors, all calibrator scans were fringe-fitted to determine the residual (time-dependent) delay and the fringe rate. 
The corrections derived from calibrators were applied to the strongest maser component, which, exhibiting a simple spatial structure consisting of a single, almost unresolved spot, was chosen to refer the visibilities of all the other maser emission channels.
The visibilities of the reference channel were fringe-fitted to find the residual fringe rate produced both by differences in atmospheric fluctuations between the calibrators and the maser, and by errors in the model used at the correlator. After correcting for the residual fringe rate, the visibilities of the reference channel were self-calibrated to remove any possible effect of spatial structure. Finally, the corrections derived from the reference channel were applied to data of all spectral channels.

Spectral channel maps were produced extending over a sky area of $(\Delta \alpha \ cos\delta \times \Delta \delta) \ 16''\times 16''$ and $ \ 2''\times 2''$  for methanol and water masers, respectively, and covering the whole velocity range where signal was visible in the total-power spectra (from  $-$7 to 5 \kms and  $-$5 to 18~km~s$^{-1}$,   at 6.7 and 22.2~GHz, respectively). H$_2$O and CH$_3$OH maser features are found to be distributed within an area of a few 0\pas1.
The CLEAN beam was an elliptical Gaussian with a  FWHM size of $ 9 \times 6 $~mas and $ 0.7 \times 0.4 $~mas, respectively at 6.7 and 22.2~GHz. In each VLBA observing epoch, the RMS noise level on the channel maps, $\sigma$, varied over a similar range of values, 7--20~mJy~beam$^{-1}$. The range of variation of the RMS noise level on the channel maps of the single-epoch EVN observations was   4--13~mJy~beam$^{-1}$. 

 The procedure to identify maser features, derive their  absolute positions, and measure their proper motions is described in detail in \citetalias{God06}. 
 In particular, for each (water and methanol) feature, the  absolute positional errors  are estimated as the sum of  the feature relative positional uncertainties (evaluated by  the weighted standard deviation of the spot positions) and the absolute positional uncertainties of the reference spot. The latter depends both on the accuracy of the absolute position of the phase-reference calibrator and on the quality of the phase-referenced map of the reference maser spot (see \citetalias{God06} for details).
The absolute positional uncertainties  of the maser features are on the order of 3 and 0.5 mas, at 6.7~GHz and 22.2~GHz, respectively (see Tables~1 and 2).


\section{Observational results}
\label{res}
 
\subsection{Sh~2-255 IR}
\begin{table*}
\centering
\caption{Parameters of water maser features detected in Sh~2-255 IR with the VLBA.}
\begin{tabular}{ccccccccccc}
\hline\hline
\multicolumn{1}{c}{Cluster} & \multicolumn{1}{c}{Feature} & \multicolumn{1}{c}{$V_{\rm L.O.S.}$} &
\multicolumn{1}{c}{$F_{\rm int}$} &  &
\multicolumn{1}{c}{$\Delta \alpha$} & \multicolumn{1}{c}{$\Delta \delta$} &
& \multicolumn{1}{c}{$V_{\rm x}$} & \multicolumn{1}{c}{$V_{\rm y}$} &
\multicolumn{1}{c}{$V_{\rm mod}$} \\
\multicolumn{1}{c}{} & & \multicolumn{1}{c}{(km s$^{-1}$)} &
\multicolumn{1}{c}{(Jy)} &   &
   \multicolumn{1}{c}{(mas)} & \multicolumn{1}{c}{(mas)} &
   & \multicolumn{1}{c}{(km s$^{-1}$)} & \multicolumn{1}{c}{(km s$^{-1}$)} &
   \multicolumn{1}{c}{((km s$^{-1}$)} \\
\hline
Aw & 1& 12.5 & 15.7 & & $-75.4 \pm 0.5$ &  $-106.6 \pm 0.2$ & & $-11 \pm 20$ & $19 \pm 9$ & $22 \pm 13$ \\
Aw & 2 & 11.9 & 0.2 & & $-80.2 \pm 0.3$ &  $-107.7 \pm 0.2$ & &  &  & \\
Aw & 3& 11.8 & 4.5 & & $-85.9 \pm 0.5$ &  $-121.0 \pm 0.3$ & & $-22 \pm 20$ & $5 \pm 11$ & $23 \pm 20$ \\
Aw & 4& 11.5 & 0.3 & & $47 \pm 0.5$ &  $-98 \pm 0.2$ & & $-13 \pm 20$ & $11 \pm 9$ & $17 \pm 17$ \\
Aw & 5 & 11.0 & 1.1 & & $-77.7 \pm 0.5$ &  $-106.5 \pm 0.2$ & & $-18 \pm 20$ & $15 \pm 9$ & $23 \pm 17$ \\
Bw & 6 & 14.4 & 0.1 & & $57.2 \pm 0.3$ &  $34.6 \pm 0.2$ & &  &  &  \\
Bw & 7& 14.3 & 0.1 & & $56.8 \pm 0.8$ &  $34.3 \pm 0.3$ & &  &  &  \\
Bw & 8 & 13.1 & 0.5 & & $56.5 \pm 0.5$ &  $0.03 \pm 0.2$ & & $-12 \pm 20$ & $41 \pm 9$ & $43 \pm 10$ \\
Bw & 9& 8.7 & 0.4 & & $96.6 \pm 0.5$ &  $-14.1 \pm 0.2$ & &  &  &  \\
Bw & 10& 8.2 & 0.4 & & $127.3 \pm 0.3$ &  $-8.9 \pm 0.2$ & &  &  &  \\
Bw & 11 & 8.1 & 0.2 & & $118.7 \pm 0.5$ &  $-114.2 \pm 0.2$ & &   &  &  \\
Bw & 12 & 8.0 & 0.09 & & $117.0 \pm 0.8$ &  $-118.4 \pm 0.3$ & &   &  &  \\
Bw & 13 & 7.6 & 0.4 & & $84.5 \pm 0.3$ &  $-21.6 \pm 0.2$ & &  &  &  \\
Bw & 14 & 7.6 & 3.9 & & $128.3 \pm 0.5$ &  $ -8.7 \pm 0.2$ & &  &  &  \\
Bw & 15 & 7.3 & 0.5 & & $103.9 \pm 0.5$ &  $-13.7 \pm 0.2$ & &  &  &  \\
Bw & 16 & 7.3 & 0.4 & & $124.9 \pm 0.5$ &  $-10.0 \pm 0.2$ & &  &  &  \\
Bw & 17& 7.3 & 2.5 & & $78.5 \pm 0.5$ &  $-19.4 \pm 0.2$ & & $12 \pm 20$ & $25 \pm 9$ & $28 \pm 12$ \\
Bw & 18 & 7.2 & 0.1 & & $85.2 \pm 0.5$ &  $-22.1 \pm 0.2$  & &  &  &  \\
Bw & 19&   6.8 & 0.2 & & $109.4 \pm 0.5$ &  $ -12.2 \pm 0.2$   & &  &  &  \\
Bw & 20 & 6.8 & 0.2 & & $83.6 \pm 0.5$ &  $-21.7 \pm 0.2$  & &  &  &  \\
Bw & 21 & 6.5 & 0.5 & & $85.9 \pm 0.5$ &  $-22.2 \pm 0.2$ & & $-12 \pm 20$ & $25 \pm 9$ & $27 \pm 12$ \\
Bw & 22 & 6.5 & 0.4 & & $128.1 \pm 0.8$ &  $-8.4 \pm 0.3$  & &  &  &  \\
Bw & 23 & 6.4 & 0.1 & & $131.0 \pm 0.5$ &  $-6.1 \pm 0.2$  & &  &  &  \\
Bw & 24 & 6.3 & 1.2 & & $115.0 \pm 0.5$ &  $-12.3 \pm 0.2$  & &  &  &  \\
Bw & 25 & 6.3 & 0.1 & & $118.8 \pm 0.3$ &  $-8.7 \pm 0.2$  & &  &  &  \\
Bw & 26 & 6.2 & 0.1 & & $128.4 \pm 0.3$ &  $-7.8 \pm 0.2$  & &  &  &  \\
Bw & 27 & 6.2 & 0.07 & & $140.3 \pm 0.5$ &  $-0.5 \pm 0.2$  & &  &  &  \\
Bw & 28 & 6.2 & 0.1 & & $217.9 \pm 0.5$ &  $54.4 \pm 0.2$  & &  &  &  \\
Bw & 29 & 6.0 & 0.8 & & $102.1 \pm 0.5$ &  $-12.6 \pm 0.2$  & &  &  &  \\
Bw & 30 & 5.8 & 0.9 & & $68.5 \pm 0.6$ &  $-2.2 \pm 0.2$ & & $13 \pm 23$ & $30 \pm 9$ & $29.9 \pm 9$ \\
Bw & 31 & 5.7 & 1.9 & & $131.6 \pm 0.5$ &  $-5.7 \pm 0.2$  & &  &  &  \\
Cw & 32 & 7.4 & 0.2 & & $-93.1 \pm 0.5$ &  $-214.4 \pm 0.2$  & &  &  &  \\
Cw & 33 & 6.6 & 0.2 & & $-133.7 \pm 0.5$ &  $-259.7 \pm 0.2$  & &  &  &  \\
Cw & 34 & 5.7 & 3.9 & & $-134.1 \pm 0.8$ &  $-260.0 \pm 0.3$  & &  &  &  \\
Cw & 35 & 5.5 & 4.9 & & $-94.7 \pm 0.8$ &  $-214.9 \pm 0.3$  & &  &  &  \\
Cw & 36 & 5.4 & 0.6 & & $-101.0 \pm 0.3$ &  $-218.7 \pm 0.2$  & &  &  &  \\
Cw & 37 & 4.8 & 0.9 & & $-142.3 \pm 0.6$ &  $-255.1 \pm 0.2$  & &  &  &  \\
Cw & 38 & 4.6 & 20.9 & & $-136.7 \pm 0.3$ &  $-261.8 \pm 0.2$  & &  &  &  \\
Cw & 39 & 4.3 & 1.1 & & $-145.5 \pm 0.5$ &  $-256.1 \pm 0.2$  & &  &  &  \\
Cw & 40 & 4.1 & 1.2 & & $-131.9 \pm 0.5$ &  $-257.4 \pm 0.2$  & &  &  &  \\
Cw & 41 & 4.1 & 0.3 & & $-93.0 \pm 0.5$ &  $-214.4 \pm 0.2$  & &  &  &  \\
Cw & 42 & 3.9 & 2.1 & & $-158.3 \pm 0.5$ &  $-249.6 \pm 0.2$ & & $ 5 \pm 20$ & $-17 \pm 10$ & $18 \pm 11$ \\
Cw & 43 & 3.8 & 0.1 & & $-146.9 \pm 0.5$ &  $-271.1 \pm 0.2$  & &  &  &  \\
Cw & 44 & 3.6 & 2.7 & & $-141.1 \pm 0.5$ &  $-254.0 \pm 0.2$ & & $ 28 \pm 20$ & $-12 \pm 9$ & $31 \pm 19$ \\
Cw & 45 & 3.4 & 0.2 & & $-131.8 \pm 0.3$ &  $-256.9 \pm 0.2$  & &  &  &  \\
Cw & 46 & 3.0 & 0.08 & & $-131.0 \pm 0.5$ &  $-255.9 \pm 0.2$  & &  &  &  \\
Cw & 47 & 2.6 & 0.1 & & $-154.7 \pm 0.3$ &  $-237.1 \pm 0.2$  & &  &  &  \\
Cw & 48 & 2.3 & 18.7 & & $-155.8 \pm 0.5$ &  $-250.2 \pm 0.2$  & &  &  &  \\
Cw & 49 & 2.2 & 6.1 & & $-155.3 \pm 0.3$ &  $-250.0 \pm 0.2$  & &  &  &  \\
Cw & 50 & 2.2 & 8.2 & & $-126.5 \pm 0.5$ &  $-242.8 \pm 0.2$ & & $ 2 \pm 20$ & $-10 \pm 9$ & $10 \pm 10$ \\
Cw & 51 & 2.1 & 0.3 & & $-155.8 \pm 0.8$ &  $-248.9 \pm 0.3$ & & $ 2 \pm 20$ & $-10 \pm 9$ & $10 \pm 10$ \\
Cw & 52 & 1.8 & 0.3 & & $-103.0 \pm 0.3$ &  $-218.5 \pm 0.2$  & &  &  &  \\
Cw & 53 & 1.7 & 1.9 & & $-167.4 \pm 0.3$ &  $-248.7 \pm 0.2$  & &  &  &  \\
Cw & 54 & 1.6 & 5.6 & & $-155.1 \pm 0.5$ &  $-250.2 \pm 0.2$  & &  &  &  \\
Cw & 55 & 1.3 & 2.3 & & $-166.0 \pm 0.5$ &  $-249.4 \pm 0.2$ & & $ -2 \pm 20$ & $-29 \pm 9$ & $29 \pm 9$ \\
Cw & 56 & 1.0 & 0.7 & & $-166.6 \pm 0.8$ &  $-249.2 \pm 0.3$   & &  &  &  \\
 $^{a}$ & 57 & 10.1 & 0.5 & & $403.8 \pm 0.8$ &  $179.9 \pm 0.3$   & &  &  &  \\
 \hline
\end{tabular}
\begin{flushleft}
{ \footnotesize  Note.-- For each identified feature,  Cols.~1  and Col.~2 give the cluster and the  feature  labels; Cols.~3 and ~4  the L.O.S. velocity  and  the integrated flux density   of the highest-intensity channel; Cols.~5 and ~6  the  positional (R.A. and Dec) offsets (with the associated errors)  evaluated with respect to the nominal maser position used at the correlator, $\alpha(J2000) = 06^h 12^m 54^s$.018,   $\delta(J2000) = 17^{\circ} 59' 23$\pas22; Cols.~7, ~8 and ~9  the projected components along the RA and Dec axes and the absolute value of the derived proper motions (with the associated errors).

(a) Isolated feature not shown in Fig.1
}
\end{flushleft}
\label{wat_vlba}
\end{table*}
\begin{figure*}
\centering
\includegraphics[angle= -90,width=\textwidth, trim = 1cm 4.5cm 0cm 0cm,clip]{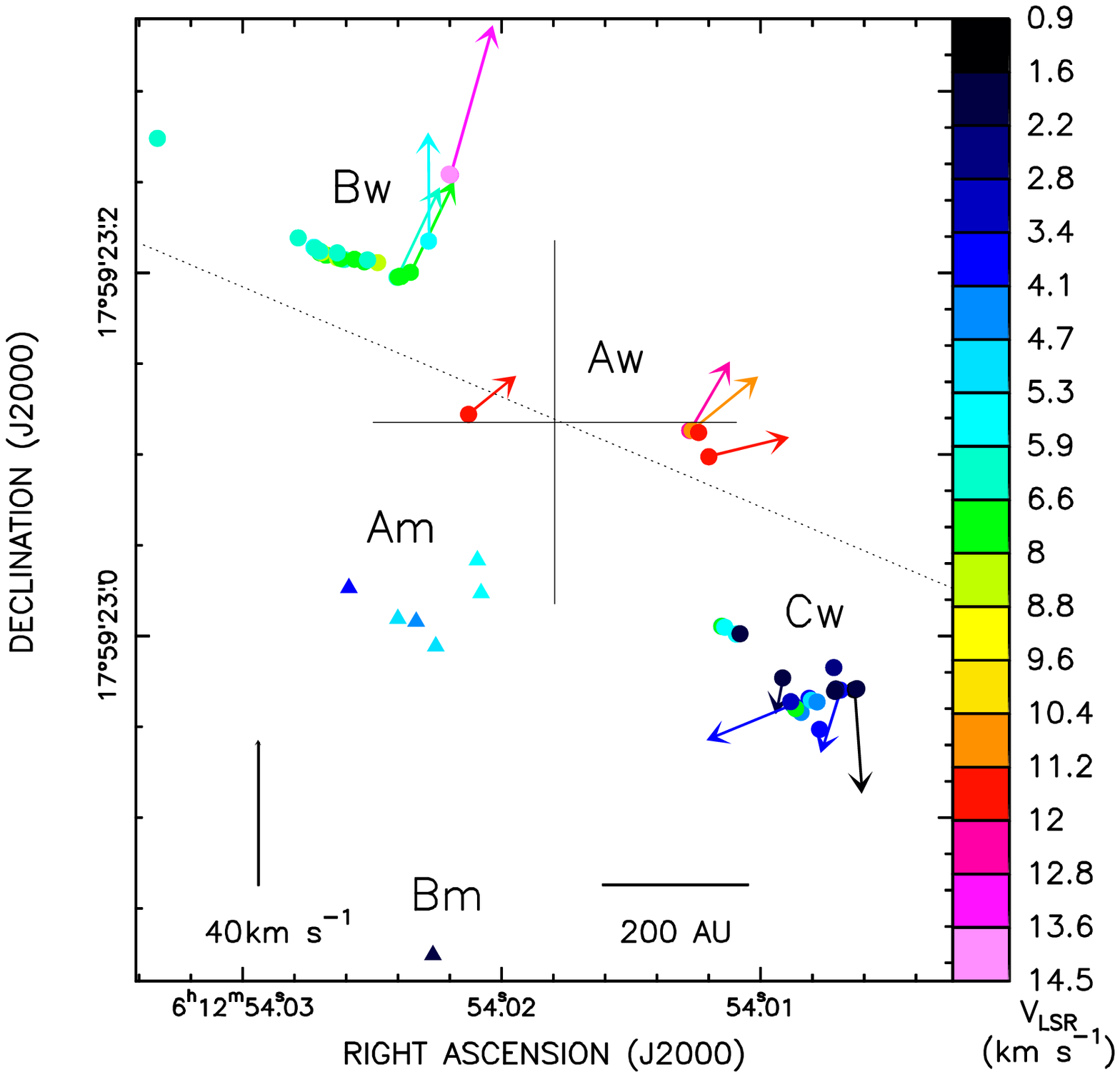}
\caption{Sh~2-255 IR. Positions and velocities of VLBA H$_2$O ({\it filled circles}) and EVN CH$_3$OH ({\it filled triangles} -- \citealt{Min00}) maser features. Different  {\em colors} denote the features' L.O.S.  velocities, according to the color scale on the right-hand side of the panel, and the {\em arrows} indicate the measured {\it absolute} proper motions of the water masers, whose amplitude scale is given in the bottom left corner of the panel;  labels Aw, Bw, and Cw identify the three clusters of VLBA water masers, whereas Am and Bm identify the two groups of methanol masers detected by \citet{Min00}. The dotted line indicates the axis of the jet observed  in the Br$\gamma $ and H$_2$ 2.12 $\mu $m lines (PA $\approx $ 67$^\circ $ -- \citealt{How97}).   The cross indicates the position uncertainty of the VLA  15~GHz continuum emission peak \citep{Ren96}. We stress that the uncertainty on the absolute position of the CH$_3$OH masers is large ($\pm$0\pas3).
}
\label{s255}
\end{figure*}

Figure~\ref{s255} shows the positions, L.O.S. velocities, and proper motions of the 22.2~GHz maser features  derived from  our multi-epoch VLBA observations, and the  positions and L.O.S. velocities of the 6.7~GHz methanol maser spots obtained from EVN observations by \citet{Min00}. 
In Fig.~\ref{s255} is also reported the positional uncertainty of the 15~GHz continuum emission observed with the VLA-B  
(beam $\approx$~0\pas5, flux density $\approx$~1.97~mJy) by \citet{Ren96}.

 The VLBA observations  
allowed us to detect a much larger number of water maser  features (57 vs 14) and  proper motions  (12 vs 3) than in our previous less sensitive EVN observations \citep{God05} . The extreme variability of the water maser emission in this source (see the discussion in Sect.~\ref{var}) may explain why only 12 maser features out of 57 were found to persist over three or four epochs, allowing  measurement of their proper motions.
The observational parameters of the  water maser features (L.O.S. velocities, flux densities, RA and Dec  positional offsets and absolute proper motions) are reported in Table~\ref{wat_vlba}.
The water maser emission is concentrated on three major clusters of features (named cluster  Aw, Bw, and Cw) plus an isolated maser feature (the last feature in Table~\ref{wat_vlba}, not shown in Fig.~\ref{s255}). Cluster Aw contains 5 features located  closer (within $\approx$100~mas) to the 15~GHz  continuum emission peak and having L.O.S. velocities red-shifted (in the range 11--12.5~\kms) with respect to the LSR velocity of the   molecular cloud  (+7.2~\kms).  Cluster Bw, comprising 26 features mostly at the systemic velocity,
is found in the north-east (NE) corner of the plotted area (Fig.~\ref{s255}), at distances $\sim$200~mas from the 15~GHz continuum emission peak. Cluster Cw includes the remaining 25 features found toward the south-west (SW) and with blue-shifted L.O.S. velocities (varying in the range 0.1--7.4~\kms). 
 In all cases, the measured {\em absolute} proper motions  have amplitudes (in the range \ 10 -- 40~km~s$^{-1}$)  large compared to the spread of  L.O.S.  velocities, covering a range of  $\ \pm7$~\kms around the  LSR  velocity of the molecular cloud (+7.2 \kms).

 The 6.7~GHz methanol maser emission appears to arise from two distinct areas, named Groups A and B by \citet{Min00}  and renamed Am and Bm by us to distinguish them from the water maser clusters.  Am and Bm are separated both in position and L.O.S. velocity.   Group Am contains 6 features spread over 70 mas (200~AU at 2.5~kpc)  in the region where the radio continuum and water maser emission are detected (Fig.~\ref{s255}). Group Am methanol spots have blue-shifted velocities, in the range 4--5.6~\kms.
  Group Bm consists of a single feature at 1.82~\kms, well separated  ($\sim$500~AU to the south)  from the radio continuum and water maser emission. 

 It is worth noting that the absolute positions of the methanol masers derived from Australia Telescope Compact Array observations are accurate within only $\sim$300~mas \citep{Min00}, so the positional association  between water and methanol masers in Sh~2-255 IR remains to be proved. 

 
\subsection{AFGL~5142}

\begin{table}
\centering
\caption{Parameters of CH$_3$OH maser features detected with the EVN in AFGL~5142.}
\begin{tabular}{ccccc}
 & & & & \\
\hline\hline
\multicolumn{1}{c}{Feature} & \multicolumn{1}{c}{$V_{\rm L.O.S.}$} &
\multicolumn{1}{c}{$F_{\rm int}$} & 
\multicolumn{1}{c}{$\Delta \alpha$} & \multicolumn{1}{c}{$\Delta \delta$} \\
\multicolumn{1}{c}{} &  \multicolumn{1}{c}{(km s$^{-1}$)} &
\multicolumn{1}{c}{(Jy)} &     \multicolumn{1}{c}{(mas)} & \multicolumn{1}{c}{(mas)}  \\
\hline
 1 & 4.9 & 0.3 &  $-9 \pm 3$ & $31 \pm 2$  \\
 2 & 4.5 & 0.3 &  $-89 \pm 3$ & $30 \pm 2$  \\
 3 & 4.4 & 0.7 &  $-2 \pm 3$ & $33 \pm 2$  \\
 4 & 4.4 & 0.7 &  $-68 \pm 3$ & $26 \pm 2$  \\
 5 & 4.1 & 1.2 &  $+9 \pm 3$ & $40 \pm 2$  \\
 6 & 4.1 & 0.9 &  $-75 \pm 3$ & $25 \pm 2$  \\
 7 & 4.0 & 0.4 &  $-63 \pm 3$ & $29 \pm 2$  \\
 8 & 3.9 & 0.4 &  $+36 \pm 3$ & $36 \pm 2$  \\
 9 & 3.8 & 0.2 &  $-7 \pm 3$ & $42 \pm 2$  \\
 10 & 3.6 & 0.9 &  $+38 \pm 4$ & $44 \pm 2$  \\
 11 & 3.6 & 0.8 &  $-70 \pm 3$ & $31 \pm 2$  \\
 12 & 3.4 & 0.8 &  $+44 \pm 3$ & $44 \pm 3$  \\
 13 & 3.0 & 0.1 &  $+24 \pm 3$ & $61 \pm 2$  \\
 14 & 2.4 & 0.5 &  $-31 \pm 3$ & $-10 \pm 2$  \\
 15 & 2.1 & 1.9 &  $+20 \pm 3$ & $74 \pm 2$  \\
 16 & 1.7 & 8.1 &  $-75 \pm 3$ & $112 \pm 3$  \\
 17 & 1.4 & 0.4 &  $+37 \pm 3$ & $103 \pm 2$  \\
 18 & 1.4 & 0.7 &  $-77 \pm 3$ & $130 \pm 3$  \\
 19 & 1.4 & 2.8 &  $-66 \pm 3$ & $128 \pm 3$  \\
 20 & 1.3 & 2.2 &  $-87 \pm 3$ & $115 \pm 3$  \\
 21 & 1.3 & 0.2 &  $+28 \pm 3$ & $98 \pm 2$  \\
 22 & 1.1 & 0.8 &  $-81 \pm 4$ & $139 \pm 4$  \\
 23 & 1.0 & 0.4 &  $-102 \pm 3$ & $117 \pm 3$  \\
 24 & 1.0 & 0.1 &  $-102 \pm 3$ & $130 \pm 2$  \\
 25 & 0.4 & 0.3 &  $-97 \pm 3$ & $106 \pm 2$  \\
 26 & -2.0 & 0.7 &  $+26 \pm 3$ & $-241 \pm 2$  \\
 27 & -6.5 & 0.2 &  $+105 \pm 3$ & $-157 \pm 2$  \\
& &  & & \\
 \hline
\end{tabular}
\begin{flushleft}
{ \footnotesize  Note.-- For each identified feature, Col.~1 gives the  label number; Cols.~2 and ~3  the L.O.S. velocity  and  the integrated flux density of the highest-intensity channel; Cols.~4 and ~5  the  positional (R.A. and Dec) offsets (and the associated errors),  evaluated with respect to the absolute position of the 22~GHz continuum peak, $\alpha(J2000) = 05^h 30^m 48^s$.02,  $\delta(J2000) = 33^{\circ} 47' 54$\pas5.
}
\end{flushleft}
\label{meth}
\end{table}
\begin{figure*}
\centering
\includegraphics[angle= -90,width=\textwidth, trim = 0cm 0cm 30cm 10cm,clip]{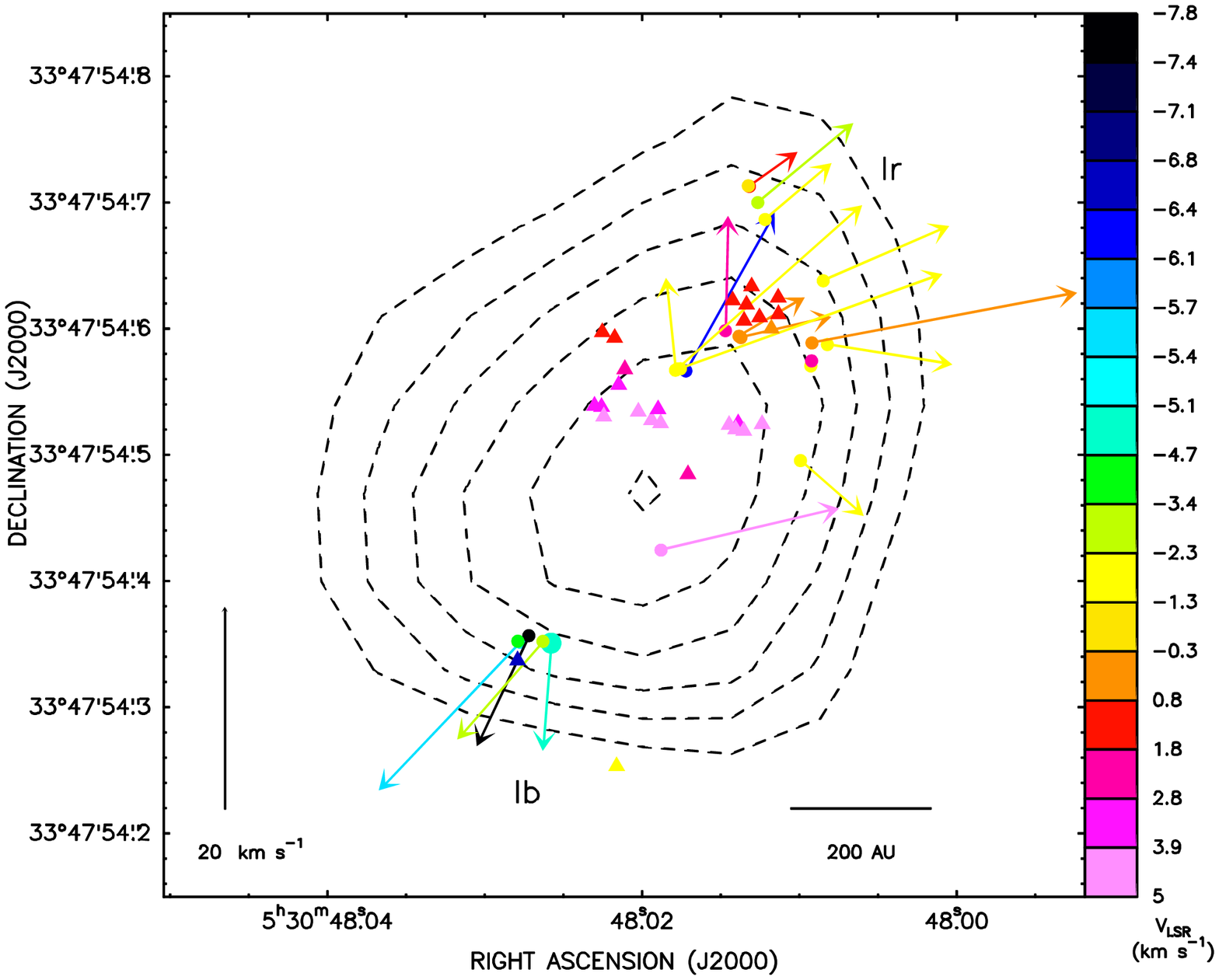}
\caption{
AFGL~5142. VLBA H$_2$O ({\it filled circles}) and EVN CH$_3$OH ({\it filled triangles})  maser features; different  {\em colors} denote the features' L.O.S.  velocities, according to the color scale on the right-hand side of the panel, and the {\em arrows} indicate the measured {\it absolute} proper motions of water masers, whose amplitude scale is given at the bottom of the panel;  the labels Ib and Ir individuate  the blue-shifted and red-shifted water maser clusters of Group I, respectively; the {\em contour} map (with contours representing 5, 6, 7, 8, and 9 times the 50~$\mu$Jy~beam$^{-1}$ RMS noise) shows the 22~GHz VLA continuum emission \citepalias{God06}.
}
\label{afgl5142}
\end{figure*}
%
 In the high-mass SFR AFGL~5142, both 6.7~GHz CH$_3$OH and 22.2~GHz H$_2$O masers are located close to free-free radio (8.4~GHz and 22~GHz) continuum emission (Fig.~\ref{afgl5142}).
Water masers have been observed at four epochs using the VLBA, which allowed the determination of the absolute positions (accurate to a few mas) and velocities of the 22~GHz maser spots \citepalias{God06}. Based on the spatial distribution and the proper motion orientation of the spots, the water maser emission appears to trace two elongated structures, indicated as Group I and Group II, oriented at quite different P.A. ($\Delta$P.A. $\ge 60^o$) (see Fig.~3 of \citetalias{God06}). Group II is associated with a north-south (NS) oriented molecular (SiO-HCO$^+$) outflow \citep{Hun99}  and will not be discussed  here.
 Masers of Group I are found closer (within $\approx$300~mas) to the 22~GHz continuum emission peak, and are divided into two spatially distinct subgroups: \ Group Ib, including features detected towards south-east (SE) of the continuum source and with L.O.S. velocities blue-shifted with respect to the  LSR  velocity of the region (--4.5~km~s$^{-1}$); \ Group Ir, located to the north-west (NW) and with red-shifted L.O.S. velocities (Fig.~\ref{afgl5142}). 

Also shown in Fig.~\ref{afgl5142}  are  the positions and L.O.S. velocities of the CH$_3$OH maser features, derived from our single-epoch EVN observations. The parameters of the  methanol maser features are reported in Table~\ref{meth}. The spatial  distribution and L.O.S. velocities of the 6.7~GHz masers resemble the bipolar structure of Group I water masers: the methanol spots found in the same area of Group Ir water masers have red-shifted L.O.S. velocities, whereas the  only blue-shifted methanol spot is observed toward  Group Ib. Some differences however are to be noted.
The methanol masers present a more compact spatial distribution,  with spots located closer to the radio continuum peak.
 The L.O.S. velocities of most of the 6.7~GHz spots fall in the range \ 1--5~km~s$^{-1}$, and are, on average, more red-shifted than those of the water masers. An interpretation for such a difference will be proposed in Sect.~\ref{afgl_dis}.

  
\section{H$_2$O  and CH$_3$OH maser variability from single-dish monitoring}
\label{var}

\begin{figure*}
\centering
\includegraphics[width=\textwidth]{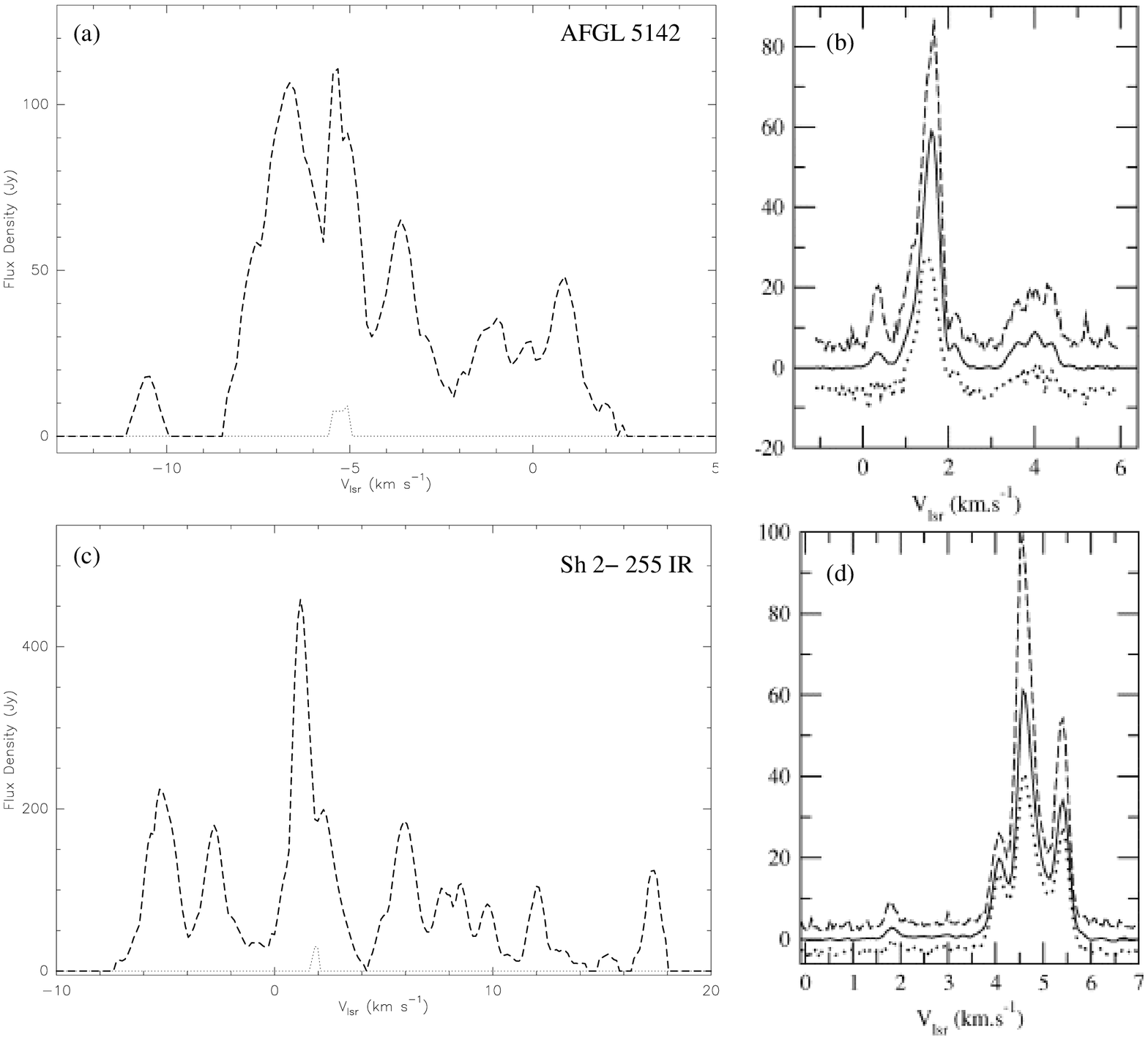}
\caption{  Range of variation in the spectrum of H$_2$O masers ({\it a} and {\it c panels}) from Medicina monitoring and CH$_3$OH masers ({\it b} and {\it d panels}) from \citet{Goe04}, in AFGL~5142 and Sh~2-255 IR respectively. The solid (for methanol), dashed and dotted lines are the averaged spectrum, the upper envelope and the lower envelope of the maser emission, respectively.}
\label{spec}
\end{figure*}

It is interesting to complement the high-angular resolution study of the  kinematics of the CH$_3$OH and  H$_2$O masers with an analysis of the time variability of the maser emission.  

 Since 1987, the Medicina 32-m antenna\footnote{The Medicina VLBI radiotelescope is operated by the INAF-Istituto di Radioastronomia.} has been used to monitor a number of galactic water masers with typical sampling intervals of 2-3 months (see \citealt{Val02} for details on the observations and data analysis). Both AFGL~5142 and Sh~2-255 IR belong to this sample, so that we could retrieve the H$_2$O spectra from the Medicina database. 
 Regarding methanol, the only published multi-year monitoring program on a large sample of class II methanol masers is that of
  \citet{Goe04}, who observed, at weekly intervals, a sample of 54 6.7-GHz methanol masers (including AFGL~5142 and Sh~2-255 IR), using the Hartebeesthoek 26-m telescope during the period 1999 January -- 2003 March. 
Following \citet{Goe04}, in order to obtain a useful visual indication of the range of variability of both maser species, Fig.~\ref{spec} reports, for both sources, the upper and lower envelopes of H$_2$O and CH$_3$OH maser emission, calculated by finding the maximum and minimum  at each velocity channel in the time interval 1999--2003. 

 Water masers are known to be highly variable (see e.g. Brand et al. 2003) and both AFGL~5142 and Sh~2-255 IR are no exception to this rule,  having spectra changing completely over the years, as demonstrated by the fact that only few spectral lines persist over three years (the lower envelopes in Figs.~3a and 3c show emission only around --5.5~\kms and 2~\kms).  Such a high variability indicates a large disturbance of the region from which water masers arise and may be caused by outflows or shock waves passing through the masing region.

In the monitored time interval, for both sources the CH$_3$OH maser emission is more stable than the water one, with the main features of methanol emitting always at the same velocities and with the relative intensities of secondary peaks changing by less than 50\%. 
 On the one hand, this higher stability  over a period of several years suggests that  methanol maser emission could arise from a less turbulent region than water masers. On the other hand, such a different temporal behavior might result from different excitation conditions of water and methanol masers. In fact, H$_2$O maser models explain the excitation by collisional pumping with H$_2$ molecules within hot ($\gtrapprox 400$~K) shocked layers of gas behind both high-velocity ($\geq$~50~km~s$^{-1}$) dissociative (J-type) \citep{Eli89} and slow ($\leq$~50~km~s$^{-1}$) non-dissociative (C-type)  \citep{Kau96} shocks, propagating in dense regions (H$_{2}$ pre-shock density $\geq 10^7$~cm$^{-3}$). In contrast, current CH$_3$OH excitation models
\citep{Cra05} predict that methanol maser emission is produced by radiative pumping in moderately warm regions ($\sim 100-200$ K) with H$_{2}$ densities $\leq 10^9$ cm$^{-3}$. Radiative pumping, assuming that the light--crossing time of the maser region is small compared with the time scale of  variability of the pumping flux, would explain the correlated variation of the intensities of diverse lines observed in the CH$_3$OH 6.7~GHz maser spectra.

In AFGL~5142 the comparison of   H$_2$O and CH$_3$OH maser spectra observed over several years evidences that  the 6.7~GHz emission steadily emerges at higher L.O.S. velocities than the water maser emission. The CH$_3$OH masers always emit in the range from --1 to +6 \kms, whereas H$_2$O masers are found in the interval from --11 to 3 \kms;  the strongest emission occurs at +2 \kms for CH$_3$OH and at --5 \kms for H$_2$O masers, respectively. 
In  Sh~2-255 IR,  the strongest CH$_3$OH emission occurs at $\sim 4.5$~\kms, close to a minimum in the H$_2$O spectra.

The different degree of time variability between water and methanol masers and, mainly, the rather clear separation in L.O.S. emission velocities are difficult to explain if the two maser species traced the same portion of gas and may suggest that, even if the two maser species traced the same kinematic structure, they  emerge from different parts of it (see Sect.~\ref{afgl_dis}).


\section{Association and kinematics of methanol and water masers in AFGL~5142}
\label{afgl_dis}
It is worth discussing the relationship between methanol and water masers in the case of AFGL~5142, where the absolute positions of both CH$_{3}$OH and H$_{2}$O maser spots have been measured with an accuracy of a few milliarcseconds. Our data reveal a {\it true} association of  both CH$_{3}$OH 6.7~GHz and H$_{2}$O 22.2~GHz masers (within a radius of a few hundreds of AU) with the same massive YSO, demonstrating that the two maser types can trace a common stage in the evolution of a forming high-mass star. 

Are methanol and water tracing the same kinematic structure? 
Based on the analysis of the single-dish spectra, we suggest that the two maser species most probably emerge from different portions of the circumstellar gas (see Sect.~\ref{var}).

H$_2$O masers clearly trace expansion from the YSO, as proved by their 3-D velocities. In \citetalias{God06} two alternative interpretations were proposed to explain the velocities of  the observed flow motion of Group I water masers. The water masers could trace the large-angle wind emerging from the atmosphere of the disk at the base of the molecular outflow observed in the HCO$^{+}$ and SiO emission{ \footnote{MHD disk--wind models predict protostellar jets to stem from the atmosphere of a Keplerian disk threaded by the circumstellar magnetic field and to collimate along the disk axis at larger distances from the YSO \citep{Kon00}.}.  Alternatively, the masers might be associated with a collimated molecular outflow emerging from  a second, massive YSO, different from that responsible for the acceleration of the HCO$^{+}$ -- SiO outflow.

 We now discuss if the new 6.7~GHz maser data comply with one of the two proposed interpretations.

If both methanol and water masers trace a collimated molecular flow,  it is difficult to explain the fact that the 6.7~GHz masers, which on the plane of the sky lie  closer to the flow center, show higher l.o.s velocities. A simple Hubble flow would require the velocities to decrease approaching the flow center, in contrast with what is observed. 


In order to test whether the 6.7~GHz maser data can be compatible with the disk-wind interpretation proposed in \citetalias{God06}, the conical flow model used in \citetalias{God06} to reproduce the motion of the 22.2~GHz masers has been applied to the 6.7~GHz masers as well. 
This model requires that the maser spots are distributed on a conical surface and move pointing away from the cone vertex. 
Using positions and velocities of both maser species, we found a best fit solution similar to that obtained using the water maser data only (see Paper I for details). 
However, this result is not conclusive since only L.O.S. velocities of the 6.7~GHz masers are measured so far, and the best fit solution is likely mainly constrained by the data of the water spots, for most of which proper motions are measured.  Only measurement of the proper motions of the 6.7~GHz maser spots can unambiguously clarify whether they are tracing the same outflowing motion as seen in the water masers.

A consistent scenario for the methanol and water maser kinematics has to explain the nature and the evolutionary stage of  the observed radio continuum source as well.  
  Based on the measured spectral properties and compact size, in \citetalias{God06} we suggest that the radio continuum emission detected toward   AFGL~5142 is originating from an hyper-compact (HC) HII region\footnote{HC HII regions are a class of very compact H II  regions, which are about 10 times smaller ($<0.01$~pc) and 100 times denser ($>10^6$~cm$^{-3}$) than  "classical" UC HII regions \citep{Kur00}.}.
 In this case, \citet{Ket03} has demonstrated that  the gravitational attraction of the star  may be sufficient  to compete with the outward force of the radiation pressure and  to balance the thermal pressure of the hot ionized gas, which then moves inward with the accretion flow \citep{Ket03}. During the earliest stages, the newly born HII region does not expand but is trapped and accretion onto the star may proceed through the HII region. 
The model of  \citet{Ket03} predicts a maximum size of the trapped HII region of the order of 100~AU for single stars or a few hundreds of AU for 
binaries and multiplets, and the minimum evolutionary timescales are \ 10$^{5}$--10$^{6}$~yr.  The predicted sizes of the trapped HII regions are
comparable with both the sky-projected separation of the  masers from the continuum peak, $\approx$400~AU, and the derived FWHM size of the 
(slightly resolved) continuum emission, $\approx$500~AU.  The predicted evolutionary timescales are long enough to make the trapped phase of an HII region much more easily observable than the following, rapid expansion phase.  A recent model by \citet{Gon05}
, studying the evolution of an HII region inside an accreting hot molecular core, indicates that the HII region expands to a size of $\sim$ 500~AU in only a few hundreds of years.

With this in mind, we consider the possibility that  the 6.7~GHz masers may trace infall rather than outflow. Since the 6.7~GHz maser spots are seen in projection against the  HC HII region and the radio continuum spectrum  indicates that  the  HC HII region is optically thick at  6.7~GHz 
(the optical depth is $\approx$ 1  at $\approx$ 8~GHz -- \citetalias{God06}), then the 6.7~GHz maser emission must be located in the foreground of the HII region. The fact that most of the spots present strongly red-shifted L.O.S. velocities suggests that they
are tracing gas infalling toward the YSO.
We tested this hypothesis with a simple model of spherical infall: $ {\bf V}({\bf r}) = -\frac{2 \ G \ M}{r^{3}} \ {\bf r} $, where \ ${\bf V}$ \ is the velocity field, ${\bf r}$ is the position vector evaluated from the YSO, $M$ is the total mass within a sphere of radius $r$ centered on the YSO, and $G$ is the gravitational constant. In order to avoid the dependence on the peculiar motion of the YSO,  the fit has been performed by comparing model and observed  velocities {\it relative} to a given maser feature, chosen as reference. With this choice, the model free parameters  are four: the sky-projected coordinates, $\alpha_{\rm s}$ (R.A.) and $\delta_{\rm s}$ (Dec), of the YSO (the sphere center); the sphere radius, $r_{s}$; and the central mass, $M_{s}$. The best-fit solution is found by minimizing the $\chi^{2}$ given by the sum of the squared differences between  the {\em relative} L.O.S.  velocities of the model and the data. The best fit parameters are: 
 $\alpha_{s}$(J2000)$ = 05^h 30^m 48^s$.018,  $\delta_{s}$(J2000)$ = 33^{\circ} 47' 54$\pas512 (coincident, within the errors, with the position of the 22.2~GHz continuum peak), 
$r_{s} = 540\pm36$~AU, and M$_{s} = 24\pm4$~M$_{\odot}$. For each fit parameter, the  error corresponds to a variation from the best fit value in correspondence of which the minimum value of  $\chi^{2}$ increases by $\sim 10$\%. Choosing a different reference maser feature for evaluating the relative velocities, the derived best fit parameters vary by  $\sim 5$\%.
The radius of the (infalling) sphere ($r_{s} = 540\pm36$~AU) is larger than the estimated size of  the HC HII regions ($r_{\rm HII} \approx 300$~AU -- \citetalias{God06}).
The value of the mass 24~M$_{\odot}$ is  a factor of two larger than the value  expected from the radio continuum emission, which requires an ionization flux of a B1 ZAMS star ($\gtrsim 10$~M$_{\odot}$ \citealt{Vac96, Pal02}).  This result is not surprising, since the mass estimated in the model is the total mass enclosed in the infalling sphere of radius $r_{\rm s}$ and thus includes also the mass of the gas.

From the fitted value of  the shell radius (540~AU) and  infall velocity (9~km~s$^{-1}$), one can derive the infall momentum rate $\dot{P}_{inf} \, = \, 1.7 \times 10^{-4}~n_{6}$~M$_{\odot}$~km~s$^{-1}$~yr$^{-1}$, where $n_{6}$ is the ambient volume density in units of \ 10$^{6}$~cm$^{-3}$.  Using a value for the far-infrared luminosity of AFGL~5142 of \ 3.8 $\times$ 10$^{3}$~L$_{\odot}$ \citep{Car90}, the radiation pressure exerted by the YSO is found to be \ 8 $\times$ 10$^{-5}$~M$_{\odot}$~km~s$^{-1}$~yr$^{-1}$, lower than the calculated infall momentum rate, which makes plausible the hypothesis that the YSO in AFGL~5142 is still accreting mass. The spherical infall model implies a mass infall rate of \ 2 $\times$ 10$^{-5}~n_{6}$~M$_{\odot}$~yr$^{-1}$,  comparable to the value estimated by \citet{Zha02} via VLA ammonia observations.
 
In conclusion, we believe that in AFGL~5142 methanol masers are probably tracing infalling gas, but proper motion measurements are needed to confirm this result. On the other hand,  the disk-wind scenario   proposed in \citetalias{God06} may  explain the expansion traced by H$_2$O masers.


\section{Water maser kinematics in Sh~2-255 IR}
\label{s255_dis}

 Sh~2-255 IR is the reddest NIR source of  the very crowded star-forming complex imaged by \citet{How97} in the NIR {\it JHK} bands.
Sh~2-255 IR   is interpreted in terms of a YSO powering an infrared  ionized jet (aligned at P.A.$\approx$67$^\circ$), detected both in the H$_{2}$ 2.122$\,\mu$m and Br$\gamma$ hydrogen recombination line emission on angular scales of 1--10 arcsec  \citep{How97}.

VLA observations in the D-configuration 
at 8, 15, 22 and 43~GHz reveal a compact radio continuum source at the position of the methanol and water masers (Minier et al., in prep.). The compact radio source is surrounded by diffuse emission that is elongated along the east-west direction. The flat spectral energy distribution measured between 8 and 43 GHz ($\sim3$~mJy) is in agreement with that of free-free emission from a ultra-compact (UC) HII region
around a ZAMS B1 star.

 Water maser emission in Sh~2-255 IR consists of three main clusters of features  distributed approximately  along a straight line,  oriented NE-SW, close to the direction of the ionized jet axis. 
This might suggest that the   water maser features are associated with the inner part of the jet/outflow system. However, in this case one would also expect the proper motions of maser spots to be mainly directed  along the jet/outflow axis. Instead, all the measured  {\em absolute} proper motions form large angles (generally $> 60^{\circ}$) with the outflow axis.
Figure~\ref{s255} shows that the water maser features in the clusters Aw and Bw move on the sky in approximately opposite directions along a line about perpendicular to the ionized jet axis (see Fig.~1).  

It is interesting to note that the spatial and velocity distribution of water masers in Sh~2-255 IR presents several similarities to that in AFGL~5142. Most of the detected water masers concentrate in two spatially separated clusters emitting at different L.O.S. velocities. The cluster separation and the range of L.O.S. velocities of water maser emission are comparable between the two sources. Also,  water masers found close to the radio continuum peak appear to trace expansion in a direction perpendicular to the axis of the jet/outflow system observed at larger ($\ge$ 1'') angular scales.  The main difference is that in  Sh~2-255 IR the sky-projected distribution of maser features is elongated in the direction of  the jet axis, rather than extending across the jet axis as observed in AFGL~5142. 

As discussed for AFGL~5142, water maser motions observed close to the YSO and directed transversally to the large scale outflow can be interpreted in the frame of the disk-wind model. The similar water maser pattern observed between AFGL~5142 and  Sh~2-255 IR lead us to speculate that the disk-wind scenario can apply to the water maser distribution in Sh~2-255 IR, as well. For this source, having measured a number of proper motions significantly lower (12 vs 23) than for AFGL~5142, no attempt is done to make more quantitative our interpretation by fitting a kinematic model.  Qualitatively, we note that the  observed spatial and velocity distribution of water masers might result from a combination of  expansion along and across the jet axis.  The disk-wind model predicts the gas motion to be directed at close angles from the disk-plane (and, hence, perpendicularly to the jet axis) in proximity to the YSO and to collimate along the disk axis at  larger distances from the YSO \citep{Kon00}. Accordingly, all maser features of the Aw cluster, found closer to the radio continuum peak and, presumably, to the YSO position, have proper motions oriented at $\approx 90^{\circ}$ from the jet axis, whilst proper motions of the Bw and Cw clusters, more distant from the YSO, appear on average to be oriented at slightly smaller angles from the jet direction.

As concerning the 6.7~GHz methanol masers observed in Sh~2-255 IR, the present data are too scarce even to attempt a qualitative interpretation of their birthplace.
We plan to conduct multi-epoch EVN observations of the 6.7~GHz methanol masers in order to determine their absolute positions and proper motions, and to ascertain their
association with the same YSO responsible for the excitation of the water maser emission. In case of finding a true association with the water masers, the planned new 6.7~GHz observations can help clarify the kinematical structures traced by both maser species toward this YSO.



\section{Conclusions}

We performed VLBA multi-epoch observations of the 22.2~GHz water masers toward the high-mass YSO Sh~2-255 IR and EVN single-epoch  observations of the 6.7~GHz methanol masers in  the high-mass SFR AFGL~5142. 

 Water maser emission in Sh~2-255 IR consists of three main clusters of features aligned along a direction close to the orientation of the molecular outflow seen  on angular scales of 1--10 arcsec, indicating that   water maser features are possibly associated with the innermost part of the jet/outflow system. However, all the measured  proper motions are not aligned along the jet axis but form large ($\geq 60^{\circ}$) angles with it. We speculate that  water masers  could trace the disk-wind emerging from the disk atmosphere at the base of the protostellar jet.

In the case of AFGL~5142,  the absolute positions of both CH$_{3}$OH and H$_{2}$O masers have been measured with an accuracy of a few milliarcseconds.
On the one hand, the interpretation of our VLBI data indicates that water masers trace expansion at the base of a molecular outflow. On the other hand, based on their peculiar positions with respect to the HII region and on their red-shifted L.O.S. velocities,  we suggest that  methanol masers in AFGL~5142  are tracing infalling rather than outflowing gas. 
Hence, in AFGL~5142 water and methanol masers, albeit associated with the same YSO, appear to trace different kinematic structures. 

\bibliographystyle{aa}
\bibliography{biblio.bib}

\begin{thebibliography}{33}
\expandafter\ifx\csname natexlab\endcsname\relax\def\natexlab#1{#1}\fi

\bibitem[{{Beuther} {et~al.}(2002){Beuther}, {Walsh}, {Schilke}, {Sridharan},
  {Menten}, \& {Wyrowski}}]{Beu02a}
{Beuther}, H., {Walsh}, A., {Schilke}, P., {et~al.} 2002, \aap, 390, 289

\bibitem[{{Carpenter} {et~al.}(1990){Carpenter}, {Snell}, \&
  {Schloerb}}]{Car90}
{Carpenter}, J.~M., {Snell}, R.~L., \& {Schloerb}, F.~P. 1990, \apj, 362, 147

\bibitem[{{Codella} \& {Moscadelli}(2000)}]{Cod00}
{Codella}, C. \& {Moscadelli}, L. 2000, \aap, 362, 723

\bibitem[{{Codella} {et~al.}(1997){Codella}, {Testi}, \& {Cesaroni}}]{Cod97}
{Codella}, C., {Testi}, L., \& {Cesaroni}, R. 1997, \aap, 325, 282

\bibitem[{{Cragg} {et~al.}(2005){Cragg}, {Sobolev}, \& {Godfrey}}]{Cra05}
{Cragg}, D.~M., {Sobolev}, A.~M., \& {Godfrey}, P.~D. 2005, \mnras, 360, 533

\bibitem[{{De Buizer}(2003)}]{Deb03}
{De Buizer}, J.~M. 2003, \mnras, 341, 277

\bibitem[{{Edris} {et~al.}(2005){Edris}, {Fuller}, {Cohen}, \& {Etoka}}]{Edr05}
{Edris}, K.~A., {Fuller}, G.~A., {Cohen}, R.~J., \& {Etoka}, S. 2005, \aap,
  434, 213

\bibitem[{{Elitzur} {et~al.}(1989){Elitzur}, {Hollenbach}, \& {McKee}}]{Eli89}
{Elitzur}, M., {Hollenbach}, D.~J., \& {McKee}, C.~F. 1989, \apj, 346, 983

\bibitem[{{Evans} {et~al.}(1977){Evans}, {Beckwith}, \& {Blair}}]{Eva77}
{Evans}, N.~J., {Beckwith}, S., \& {Blair}, G.~N. 1977, \apj, 217, 448

\bibitem[{{Goddi} \& {Moscadelli}(2006)}]{God06}
{Goddi}, C. \& {Moscadelli}, L. 2006, \aap, 447, 577

\bibitem[{{Goddi} {et~al.}(2005){Goddi}, {Moscadelli}, {Alef}, {Tarchi},
  {Brand}, \& {Pani}}]{God05}
{Goddi}, C., {Moscadelli}, L., {Alef}, W., {et~al.} 2005, \aap, 432, 161

\bibitem[{{Goedhart} {et~al.}(2004){Goedhart}, {Gaylard}, \& {van der
  Walt}}]{Goe04}
{Goedhart}, S., {Gaylard}, M.~J., \& {van der Walt}, D.~J. 2004, \mnras, 355,
  553

\bibitem[{{Gonz{\'a}lez-Avil{\'e}s} {et~al.}(2005){Gonz{\'a}lez-Avil{\'e}s},
  {Lizano}, \& {Raga}}]{Gon05}
{Gonz{\'a}lez-Avil{\'e}s}, M., {Lizano}, S., \& {Raga}, A.~C. 2005, \apj, 621,
  359

\bibitem[{{Howard} {et~al.}(1997){Howard}, {Pipher}, \& {Forrest}}]{How97}
{Howard}, E.~M., {Pipher}, J.~L., \& {Forrest}, W.~J. 1997, \apj, 481, 327

\bibitem[{{Hunter} {et~al.}(1999){Hunter}, {Testi}, {Zhang}, \&
  {Sridharan}}]{Hun99}
{Hunter}, T.~R., {Testi}, L., {Zhang}, Q., \& {Sridharan}, T.~K. 1999, \aj,
  118, 477

\bibitem[{{Kaufman} \& {Neufeld}(1996)}]{Kau96}
{Kaufman}, M.~J. \& {Neufeld}, D.~A. 1996, \apj, 456, 250

\bibitem[{{Keto}(2003)}]{Ket03}
{Keto}, E. 2003, \apj, 599, 1196

\bibitem[{{Konigl} \& {Pudritz}(2000)}]{Kon00}
{Konigl}, A. \& {Pudritz}, R.~E. 2000, Protostars and Planets IV, 759

\bibitem[{{Kurtz} {et~al.}(2000){Kurtz}, {Cesaroni}, {Churchwell}, {Hofner}, \&
  {Walmsley}}]{Kur00}
{Kurtz}, S., {Cesaroni}, R., {Churchwell}, E., {Hofner}, P., \& {Walmsley},
  C.~M. 2000, Protostars and Planets IV, 299

\bibitem[{{Mezger} {et~al.}(1990){Mezger}, {Zylka}, \& {Wink}}]{Mez90}
{Mezger}, P.~G., {Zylka}, R., \& {Wink}, J.~E. 1990, \aap, 228, 95

\bibitem[{{Minier} {et~al.}(2000){Minier}, {Booth}, \& {Conway}}]{Min00}
{Minier}, V., {Booth}, R.~S., \& {Conway}, J.~E. 2000, \aap, 362, 1093

\bibitem[{{Minier} {et~al.}(2001){Minier}, {Conway}, \& {Booth}}]{Min01}
{Minier}, V., {Conway}, J.~E., \& {Booth}, R.~S. 2001, \aap, 369, 278

\bibitem[{{Moscadelli} {et~al.}(2005){Moscadelli}, {Cesaroni}, \&
  {Rioja}}]{Mos05}
{Moscadelli}, L., {Cesaroni}, R., \& {Rioja}, M.~J. 2005, \aap, 438, 889

\bibitem[{{Norris} {et~al.}(1998){Norris}, {Byleveld}, {Diamond}, {Ellingsen},
  {Ferris}, {Gough}, {Kesteven}, {McCulloch}, {Phillips}, {Reynolds},
  {Tzioumis}, {Takahashi}, {Troup}, \& {Wellington}}]{Nor98}
{Norris}, R.~P., {Byleveld}, S.~E., {Diamond}, P.~J., {et~al.} 1998, \apj, 508,
  275

\bibitem[{{Palla} {et~al.}(2002){Palla}, {Zinnecker}, {Maeder}, \&
  {Meynet}}]{Pal02}
{Palla}, F., {Zinnecker}, H., {Maeder}, A., \& {Meynet}, G., eds. 2002,
  {Physics of star formation in galaxies}

\bibitem[{{Pestalozzi} {et~al.}(2004){Pestalozzi}, {Elitzur}, {Conway}, \&
  {Booth}}]{Pes04}
{Pestalozzi}, M.~R., {Elitzur}, M., {Conway}, J.~E., \& {Booth}, R.~S. 2004,
  \apjl, 606, L173

\bibitem[{{Rengarajan} \& {Ho}(1996)}]{Ren96}
{Rengarajan}, T.~N. \& {Ho}, P.~T.~P. 1996, \apj, 465, 363

\bibitem[{Snell {et~al.}(1988)Snell, Huang, Dickman, \& Claussen}]{Sne88}
Snell, R.~L., Huang, Y.-L., Dickman, R.~L., \& Claussen, M.~J. 1988, ApJ, 325,
  853

\bibitem[{{Sridharan} {et~al.}(2002){Sridharan}, {Beuther}, {Schilke},
  {Menten}, \& {Wyrowski}}]{Sri02}
{Sridharan}, T.~K., {Beuther}, H., {Schilke}, P., {Menten}, K.~M., \&
  {Wyrowski}, F. 2002, \apj, 566, 931

\bibitem[{{Vacca} {et~al.}(1996){Vacca}, {Garmany}, \& {Shull}}]{Vac96}
{Vacca}, W.~D., {Garmany}, C.~D., \& {Shull}, J.~M. 1996, \apj, 460, 914

\bibitem[{{Valdettaro} {et~al.}(2002){Valdettaro}, {Palla}, {Brand},
  {Cesaroni}, {Comoretto}, {Felli}, \& {Palagi}}]{Val02}
{Valdettaro}, R., {Palla}, F., {Brand}, J., {et~al.} 2002, \aap, 383, 244

\bibitem[{{Walsh} {et~al.}(1998){Walsh}, {Burton}, {Hyland}, \&
  {Robinson}}]{Wal98}
{Walsh}, A.~J., {Burton}, M.~G., {Hyland}, A.~R., \& {Robinson}, G. 1998,
  \mnras, 301, 640

\bibitem[{Zhang {et~al.}(2002)Zhang, Hunter, Sridharan, \& Ho}]{Zha02}
Zhang, Q., Hunter, T., Sridharan, T., \& Ho, P. 2002, ApJ, 566, 982

\end{thebibliography}

\end{document}